# Andreev reflection versus Coulomb blockade in hybrid semiconductor nanowire devices

Yong-Joo Doh,[1] Silvano De Franceschi,[2,3,*] Erik P. A. M. Bakkers,[4]
and Leo P. Kouwenhoven[1]

[1]Kavli Institute of Nanoscience, Delft University of Technology, PO Box 5046, 2600
GA Delft, Netherlands
[2]CEA, INAC/SPSMS/LaTEQS, 17 rue des Martyrs, 38054 Grenoble, France
[3]CNR-INFM, Laboratorio Nazionale TASC, S.S. 14, Km 163.5, 34012 Trieste, Italy
[4]Philips Research Laboratories, Professor Holstlaan 4, 5656 AA Eindhoven,
Netherlands

[*] Corresponding author. E-mail: silvano.defranceschi@cea.fr

**Semiconductor nanowires provide promising low-dimensional systems for the study of quantum transport phenomena in combination with superconductivity. Here we investigate the competition between the Coulomb blockade effect, Andreev reflection, and quantum interference, in InAs and InP nanowires connected to aluminum-based superconducting electrodes. We compare three limiting cases depending on the tunnel coupling strength and the characteristic Coulomb interaction energy. For weak coupling and large charging energy, negative differential conductance is observed as a direct consequence of the BCS density of states in the leads. For intermediate coupling and charging energy smaller than the superconducting gap, the current-voltage characteristic is dominated by Andreev reflection and Coulomb blockade produces an effect only near zero bias. For almost ideal contact transparencies and negligible charging energy, we observe universal conductance fluctuations whose amplitude is enhanced due to Andreev reflection at the contacts.**

In recent years semiconductor nanowires (NW's) have attracted a lot of attention owing to their potential for applications in the domain of nanoscale sensors, electronics and photonics.[1] Simultaneously, the highly controlled synthesis of homo- and hetero-structured NW's has provided a new platform to study fundamental quantum phenomena and develop novel quantum devices[2]. Resonant tunneling[3] and Coulomb blockade[4-6] behavior have been observed in the low-temperature electronic transport[7] of individual NW's. Also, NW's have been efficiently connected to conventional superconductors (SC's) leading to the observation of gate-tunable supercurrents[8, 9], π-junction behavior[10], and Kondo-enhanced Andreev tunneling.[11] While in all these experiments the NW-SC tunnel coupling ranged between moderate and strong, relatively less attention has been paid to the weak coupling limit, where the characteristic energy scales associated with the Coulomb blockade effect and with size quantization largely exceed the superconducting energy gap. This regime was addressed for the first time by Ralph *et al.*[12] who could obtain quantum dots weakly coupled to SC's by embedding metallic nanoparticles in a thin oxide layer sandwiched between two SC contacts. In this pioneering work, negative differential conductance (NDC) was observed as a hallmark of the BCS density of states (DOS) in the source/drain electrodes.[13, 14] Further studies of the dependence on particle size or tunnel coupling, however, were hindered by experimental difficulties in the device fabrication process. Other more recent attempts using either carbon nanotubes[15] or self-assembled InAs nanocrystals[16] lacked a clear evidence of NDC.

In this Letter we report an extensive study of SC-NW-SC hybrid systems embracing a wide range of contact transparencies, from very low to almost ideal ones. The observed transport properties are determined by an interplay of different mesoscopic phenomena with competing energy scales including: *electron-electron Coulomb interactions*, characterized by a charging energy $E_c = e^2/C_\Sigma$ where $e$ is the electron charge and $C_\Sigma$ the total capacitance of the NW; *Andreev reflection*, determined by the energy gap, $\Delta$, of the superconducting electrodes; and *phase-coherent diffusive transport*, related to the Thouless energy, $E_{Th} = h/\tau_D$, where $h$ is Planck's constant and $\tau_D$ the characteristic time an electron takes to diffuse along the all NW length. We consider three limiting cases corresponding to clearly different coupling strengths between the NW and the SC contacts: 1) a weak tunnel coupling regime, with $k_B T \ll \Delta \ll E_c$ ($T$ is temperature and $k_B$ Boltzmann's constant), in which transport is dominated by the Coulomb blockade phenomenon, and the superconducting nature of the electrodes manifests itself through its effect on their quasiparticle density of states (DOS); 2) an intermediate coupling regime, with $k_BT \ll E_c \ll \Delta$, in which electronic transport is dominated by Andreev reflection and Coulomb blockade emerges only at low bias voltage; 3) a strong coupling regime, with $E_c \ll k_BT \ll \Delta$, $E_{Th}$, characterized by a combination of universal conductance fluctuations and Andreev reflection in the absence of Coulomb blockade.

The NW's employed for this work were synthesized by a vapor-liquid-solid process, in which the semiconducting materials were supplied via laser ablation of *n*-doped InP or InAs targets. The NW growth was catalyzed by Au nanoparticles pre-deposited on a Si substrate. After growth, the NW's were dispersed in a chlorobenzene solution and deposited on a $p^+$ Si substrate with a 250-nm-thick SiO$_2$ overlayer. The SC contact electrodes were fabricated by e-beam lithography followed by the e-beam deposition of a Ti(10 nm)/Al(120 nm) bilayer. The NW surface was treated with buffered hydrofluoric acid for 20 s prior to metal evaporation. All low-temperature transport measurements were carried out in a dilution refrigerator equipped with accurately filtered wiring and low-noise electronics. Further details on NW growth, device fabrication, and measurement setup can be found in earlier publications[4, 8, 17]. The normal state resistance $R_N$, which we define as the dynamic resistance, $dV/dI$, at a bias voltage, $V$, of 1 mV and $T = 2$ K, was found to vary between ~1 to ~$10^3$ k$\Omega$ depending mostly on NW composition and diameter (see Table. S1 and Fig. S1 in the Supplementary Information).

We consider first the device with the largest resistance, fabricated from a 36-nm-diam InP NW (see Fig. 1a). The device, labeled as **D1**, has $R_N/R_Q = 36$, where $R_Q = h/e^2 \sim 25.8$ k$\Omega$ is the resistance quantum. The NW channel forms a small quantum dot whose low-temperature transport properties are dominated by the Coulomb blockade effect over the whole experimental range, as denoted by the presence of characteristic quasi-periodic peaks[18] in the linear conductance, $G$ (see Fig. 1b). Such Coulomb peaks were measured at $T = 20$ mK under a perpendicular magnetic field of 0.1 T, large enough to suppress superconductivity in the Ti/Al electrodes. From the average peak-to-peak separation, $\delta V_g \sim 0.1$ V, we estimate a gate capacitance $C_g = e/\delta V_g \sim 2$ aF. Based on a simplified model – we use $C_g = 2\pi\varepsilon_0\varepsilon L_0/\ln(2h/r)$, where $\varepsilon_0$ is the vacuum permittivity, $\varepsilon \sim 2.5$ an effective dielectric constant accounting for presence of a SiO$_2$ layer under the NW and vacuum above it, $L_0$ the length of the NW quantum dot, $h = 250$ nm the thickness of SiO$_2$ layer, and $r$ the NW radius – this $C_g$ value corresponds to a dot length, $L_0 \sim 50$ nm, smaller than the source-drain distance $L = 90$ nm. This discrepancy can be ascribed to the electrostatic screening effect of the source and drain electrodes in such a short-channel device.[6]

More quantitative information on the NW quantum dot can be obtained from a measurement of the dynamic conductance, $dI/dV$, as a function of $V_g$, and $V$. The resulting stability diagram for normal leads is shown in Fig. 1c. The height of the Coulomb diamonds yields $E_c \sim 0.7$ meV, corresponding to a total capacitance of the dot $C_\Sigma = C_s + C_d + C_g \sim 230$ aF, where $C_s$ and $C_d$ are the capacitances of the source and drain tunnel junctions, respectively. Given a ratio $C_d/C_s \sim 1.6$, obtained directly from the slopes of the diamond edges[19], we estimate $C_s = 88$ aF and $C_d = 140$ aF. The discrete energy spectrum of the NW quantum dot is revealed by the presence of additional lines parallel to the diamond edges. These lines are due to single-electron tunneling via the excited states of the dot, indicating a characteristic level spacing $\delta E = 0.2 - 0.3$ meV[7].

The stability diagram changes dramatically as the magnetic field is turned off to revive superconductivity in the Ti/Al leads. As shown in Fig. 1(d) the Coulomb diamonds are split apart along the $V$-axis leading to the appearance of an almost "currentless window" around zero bias[15]. This window reflects the opening of superconducting gaps in the leads. In the case of ideal gaps with no quasiparticle states, current flow should be blocked all over the bias range between $-2\Delta/e$ and $2\Delta/e$. Figure 1(d), however, displays two overlapping windows of suppressed conduction bound by $V = \pm 0.13$ and $\pm 0.26$ mV, corresponding to $\pm\Delta/e$ and $\pm 2\Delta/e$, respectively. Moreover, some weak conductance peaks can still be seen around zero bias at the charge-degeneracy points as a reminiscence of the diamond edges in the normal state. We ascribe these observations to the existence of quasiparticle states in the superconducting gap of the leads, which can be due to the granularity of the metallic contacts[20] and/or a non-zero DOS in the NW sections adjacent to the quantum dot.

For $|V| > \Delta/e$, the Coulomb peaks associated with the source lead (lines from bottom-left to top-right in Figs. 1c,d) exhibit a sensible enhancement with respect to the normal case and they are followed by broader NDC lines (dark blue regions). We ascribe this combination of enhanced $dI/dV$ peaks and NDC to a BCS-like DOS in the source electrode. The single-electron tunneling current has a maximum when the resonant dot level is aligned with the DOS peak at the edge of the superconducting gap, and it decreases at larger bias due to a reduction in the available DOS leading to NDC. This contrasts with the usual case of a quantum dot connected to normal leads where the available DOS is approximately constant and a quantum-dot level entering the bias window produces a step-like increase in the current, i.e. a $dI/dV$ peak with no NDC. (We note that another example of NDC was reported for NW resonant tunneling diodes and it was explained as an effect of the finite bandwidth of conduction electrons in the NW source lead.[3])

To illustrate the development of NDC resulting from the onset of superconductivity we show in Fig. 1e a set of $I(V)$ traces for a fixed gate voltage, $V_g$ = -0.26 V, and different magnetic fields, from $B$ = 20 mT to $B$ = 0. In the normal state ($B$ = 20 mT) the $I(V)$ exhibits the characteristic current steps expected for a constant DOS in the leads. Upon decreasing $B$, the current steps evolve into peaks leading to NDC. The gradual shift in the peak position towards larger $|V|$ reflects the increase in the superconducting energy gap. The quantitative relation between this shift, $\delta V(B)$, and $\Delta(B)$ can be obtained by taking into account the bias voltage division between the two tunnel junctions at the source and drain contacts. It can be shown that $\delta V = (C_s+C_d)\Delta/eC_s$ (or $(C_s+C_d)\Delta/eC_d$), when the initial tunneling step occurs through the drain (or source) junction[12]. Unlike in the experiment by Ralph *et al.*[12], here all dot capacitances can be directly extracted from the Coulomb diamonds in Fig. 1c without the need for multi-parameter fits. From $C_d/C_s$ ~ 1.6, we obtain $|\delta V|$ = 1.6$\Delta$ and 2.6$\Delta$ for positive and negative $V$, respectively. Between $B$ = 20 mT and 0 mT , we measure $\delta V$ = 0.22 mV for $V > 0$, corresponding to $\Delta$ = 0.14 meV, a value that is consistent with previous experiments[8,22]. For $V < 0$, the first resonance shows an anomalous shift of only 0.05mV. In Ref. 12, similar anomalous shifts were interpreted as a possible result of additional single-electron tunneling resonances via dynamically populated excited states. Here the anomaly has a simpler origin that can be understood by examining the $B$-evolution of the single-electron tunneling edges. As shown in the lower inset of Fig. 1e and further clarified by the schematic stability diagram in the upper inset, the observed small shift originates from a resonance with the source and not with drain as one might expect.

We now consider the intermediate coupling regime with $R_N/R_Q$ ~ 1 and $k_B T \ll E_c \ll \Delta$. Figure 2(a) shows a set of $dI/dV(V)$ characteristics at different temperatures for a device, **D2,** with channel length $L$ = 220 nm, diameter $\phi$ = 44 nm and $R_N/R_Q$ = 0.77. Below the superconducting transition temperature, $T_c$ ~ 1.3 K, a conductance dip develops at low bias together with two peaks at $V = \pm 2\Delta/e$. This behavior is typical of superconductor-insulator-normal metal-insulator-superconductor (SINIS) junctions in which conduction for $|V| < 2\Delta/e$ relies on Andreev reflection[21]. The fact that $dI/dV$ in this Andreev-dominated range is suppressed with respect to the normal-state $dI/dV$ (measured at high bias or for $T > T_c$) is due to a low transparency of the interfaces between the NW and the SC electrodes.[22]

An additional conductance dip appears at zero bias with a half width $V_{HW} = 28$ μV at $T = 20$ mK. The depth of this dip is found to oscillate periodically with $V_g$ as denoted by the oscillations in the linear conductance shown in Fig. 2(b). At $T = 20$ mK, these oscillations acquire the form of regularly spaced conductance peaks that can be fitted very well to a "classical" Coulomb blockade model[18], $G/G_{max} \sim \cosh^{-2}[(e(C_g/C_\Sigma)|V_{g,peak} - V_g|)/2.5k_BT]$, with $G_{max}$ and $V_{g,peak}$ being the peak height and position. From such a fit, with a gate capacitance $C_g = 7.5$ aF, derived directly from the oscillation period $\delta V_g = 21$ mV, we obtain a charging energy $E_c = e^2/C_\Sigma \sim 13\ k_BT$. Since $T$ should be replaced by the effective electron temperature, which we estimate to be between 25 and 40 mK, we obtain $E_c = 20 - 45$ μeV consistent with the half width of the dips. With increasing $T$, the Coulomb oscillations are suppressed above 0.3 K, i.e. for $k_BT > E_c$. In contrast to the case of **D1**, $E_c << \Delta$ implies that many conduction channels become accessible in the NW quantum dot when $E_c << eV < \Delta$. This leads to the complete disappearance of the Coulomb-blockade effect in this bias range where transport follows the expectation for a non-interacting SINIS junction.

To further explore the non-interacting limit down to very low energy scales we have studied NW devices with $R_N << R_Q$, and an always negligible Coulomb energy ($E_c << k_BT$). Although $R_N/R_Q << 1$ could be obtained with InP NW's of relatively large diameter (see e.g. device **D3** in Table S1 of the Supplementary Information), this low-resistance regime was more efficiently achieved with InAs NW's owing to the known ability of this material to form Schottky-barrier-free contacts with metals. A typical magnetoconductance $G(B)$ curve at $T = 30$ mK is shown in the inset of Fig. 3a for device **D5**, fabricated from a 80-nm-diam InAs NW ($L = 110$ nm). The magnetic field was applied perpendicular to the nanowire axis. The NW conductance shows reproducible aperiodic fluctuations with a root-mean square amplitude, rms($G_B$) $\equiv$ $<(G(B)-<G(B)>)^2>^{1/2} = 0.29$ and 0.30 $e^2/h$, for devices **D4** and **D5**, respectively (here angular brackets indicate an average over magnetic field between -5 and 5 T with the exclusion of a 1-Tesla-wide window around zero field where G exhibits an overshoot due to superconductivity as well as possible variations related to weak (anti)localization). These so-called universal conductance fluctuations (UCF) are a mesoscopic effect arising from the quantum interference of multiply scattered electronic wave packets[23]. UCF are known to appear in diffusive conductors when their size is smaller or comparable to the characteristic phase-coherence length, $L_\phi$.

An estimate of $L_\phi$ can be obtained from the autocorrelation function $F(\Delta B) = <G(B)G(B+\Delta B)> - <G(B)>^2$ with $\Delta B$ a lag parameter in magnetic field[24, 25]. $F(\Delta B)$ is expected to have a peak at $\Delta B = 0$. The half-width at half height of this peak corresponds to a magnetic correlation length, $B_c$ (conductance values measured at two different fields are uncorrelated when the field difference is much larger than $B_c$). Figure 3b shows the positive side ($\Delta B > 0$) of the autocorrelation function. From the data obtained in perpendicular magnetic field (open dots) we find $B_c = 0.21$ T (**D4**) and 0.18 T (**D5**). According to a theoretical model for quasi-one-dimensional conductors[26], $B_c = 0.42\ \Phi_0/(wL_\phi)$, where $\Phi_0 = h/e$ is the one-electron flux quantum and $w$ a width corresponding to the NW diameter. From $B_c \approx 0.2$ T and $w = 80$ nm we find $L_\phi \approx 100$ nm. This value is smaller than the one obtained from weak localization/antilocalization measurements in similar NW's [27]. Similar $F(\Delta B)$ curves are found for $B$ applied parallel to the NW axis (solid lines in Fig. 3a), in contrast with previous results for multi-walled carbon nanotubes[28]. We argue that the seemingly weak dependence of $B_c$ on the field direction reflects the fact that $L_\phi$ is close to the NW diameter.

UCF have been observed also at $B = 0$ T, using the back-gate voltage as varying parameter ($V_g$ changes the electron density as well as the impurity charge configuration in the nanowire thereby affecting the quantum interference of scattered electrons). The inset of Fig. 3b shows bias-dependent $dI/dV(V_g)$ curves for **D4**. After subtracting the average background conductance obtained from a fit to a second order polynomial function, we find rms($G_g$) = $0.48 \pm 0.02\ e^2/h$ at $V > 2\Delta/e \sim 0.23$ mV. In this large-bias regime, superconducting correlations are expected to vanish similar to the normal-state case discussed above. In the low-bias regime, $V < 2\Delta/e$, the fluctuation amplitude increases substantially (up to a factor 1.6) while its pattern is preserved. We interpret this enhancement of rms($G_g$) below the superconducting gap energy as the result of the correlated interference of additional hole-like quasiparticles generated in the nanowire by Andreev reflection processes at the contacts, and propagating along the time-reversed path of their electron-like counterparts. An alternative possibility to probe the contribution of Andreev reflection processes is to apply a constant magnetic field that suppresses superconductivity. In the normal state regime at $B = 0.1$ T, we find rms($G_g$)~$0.47\ e^2/h$ at $V$=0, which is close to the high-bias value at $B = 0$. This supports our interpretation in terms of an interplay between UCF and Andreev reflection.

In summary, we have reported the observation of different transport regimes in InP and InAs NW's coupled to superconducting electrodes. The relative weight of Coulomb interactions and Andreev reflection was found to vary from device to device depending on the tunnel coupling strength. Additionally, superconductivity was selectively switched on and off with a moderate magnetic field. In the weak coupling limit, transport occurs via single-electron tunneling due to Coulomb blockade, and negative differential conductance is observed as a result of BCS singularities in the density of states of the electrodes. For strong tunnel coupling, transport is dominated by Andreev reflection leading to an enhancement of the conductance and its mesoscopic fluctuations. In the intermediate regime, we considered for the first time the case of a Coulomb charging energy much smaller than the superconducting gap. This energy scale separation results in three clearly distinct regimes: $|V| < E_c$, $E_c < |V| < 2\Delta$, and $|V| > 2\Delta$, fully dominated by Coulomb blockade, Andreev reflection, and non-interacting Ohmic-like transport, respectively. This work emphasizes the potential of semiconductor nanowires as versatile material systems for the investigation of complex fundamental problems involving various competing phenomena such as superconductivity. Along this line several interesting problems could be addressed in the near future starting from the interplay between the superconducting proximity effect and spin-dependent phenomena such the Kondo effect and spin-orbit coupling.

,

**Acknowledgments**
We thank J. A. van Dam, A. Roest, Yu. Nazarov, C. W. J. Beenakker and Kicheon Kang for helpful discussions. We acknowledge financial support from the EU through the HYSWITCH project, from CNR through the EUROCORES FoNE programme (contract No. ERAS-CT-2003-980409), and from the National Creative Research Initiative Project (R16-2004-004-01001-0) of the Korea Science and Engineering Foundations..

**Figure Captions**

**Figure 1.** (a) Scanning electron micrograph of device **D1**, with inter-electrode distance $L$ = 90 nm and nanowire diameter $\phi$ = 36 nm. Scale bar: 500 nm. (b) Back-gate voltage, $V_g$, dependence of device **D1** linear conductance, $G$, at $T$ = 20 mK under perpendicular magnetic field $B$ = 100 mT (normal-state contacts) 0. (c) Color plot of differential conductance, $dI/dV$, as a function of bias voltage, $V$, and $V_g$ in the normal state ($B$ = 31 mT) and (d) in the superconducting state ($B$ = 0 mT). Dark blue regions in (d) correspond to $dI/dV$ < 0. (e) Current-voltage $I(V)$ characteristics at $T$ = 20 mK, $V_g$ = -258 mV, and $B$ = 20 (black), 16 (red), 12 (green), 8 (blue), 4 (cyan), and 0 (magenta) mT. Lower inset: color plot of $dI/dV$ vs ($V$,$B$) for the same $T$ and $V_g$. Below the critical magnetic field of the contacts (~15 mT), the dI/dV peaks associated with the onset of single-electron tunneling shift according to the opening of a superconducting gap in the contacts (see dashed lines). Negative dI/dV is observed near $B$ = 0 only for resonances associated with the source contact (yellow and green lines). The first drain-related resonance (blue line) exhibits a maximum $V$-shift of 0.37 mV (consistent with $\Delta$ = 0.14 meV) but no negative $dI/dV$, suggesting a pronounced broadening in the BCS-like density of states of the drain contact. Upper inset: schematic view of stability diagram for the case of normal ($B$ = 31 mT, thin solid lines) and superconducting ($B$ = 0, thick solid lines) contacts. The vertical dashed line corresponds to $V_g$ = -258 mV, and the colored circles indicate the observed $dI/dV$ peaks at B = 0.

**Figure 2.** (a) Temperature dependence of differential conductance $dI/dV(V)$ for device **D2** with $B$ = 0 mT. From top to bottom: $T$ = 1.67, 1.30, 1.12, 0.84, 0.46 and 0.02 K. $dI/dV$ peak occurs at $V$ = ± 370 μV at the lowest temperature of $T$ = 20 mK. (b) Linear conductance $G(V_g)$ curve at $T$ = 20 mK with $B$ = 0 mT. (c) $T$ dependence of $G(V_g)$ curve with $T$ = 20 (open circle), 70 (closed circle), 140 (open square), 200 (closed triangle), 250 (open triangle), and 310 mK (closed square). The solid line is a fit to the Coulomb

oscillations in the classical regime with $k_{B}T/(e^2/C) = 0.076$ (see text).

**Figure 3.** (a) Autocorrelation functions $F(\Delta B)$, extracted from the magnetoconductance $G(B)$ curves at $V_{g} = 0$ V, for **D4** (red) and **D5** (black) with perpendicular (line) and parallel (circle) magnetic field $B$ to the nanowire axis. The source-drain spacing is $L = 440$ (**D4**) and 110 (**D5**) nm, respectively, and $\phi = 83$ nm. Inset: $G(B)$ curve, obtained from **D5**, with perpendicular $B$ at $T = 30$ mK. The overshoot of $G$ near $B = 0$ T is due to a supercurrent through the nanowire. (b) $V$-dependent rms($G_{g}$) for device **D4** (filled circle) and **D6** (square) with $B = 0$ T. For $V = 0$, rms($G_{g}$) = 0.47 $e^2/h$ (open circle) is obtained from **D4** in the normal state with $B = 0.1$ T. Inset: $V_{g}$-dependent dynamic conductance curves for **D4** taken at $T = 22$ mK and different bias voltages $V = 0.11$ (black), 0.24 (red), 0.44 (green), and 0.65 (blue) mV.

**Supplementary Information**

**Table. S1** Basic characteristics of the 6 representative devices discussed in the main text. $\phi$ is the nanowire diameter, $L$ the spacing between source and drain electrodes, $R_N$ the normal-state device resistance in units of $h/e^2$ (= 25.8 k$\Omega$).

| No. | nanowire | $\phi$ (nm) | $L$ (nm) | $R_N$ ($h/e^2$) |
|---|---|---|---|---|
| **D1** | InP | 36 | 90 | 36 |
| **D2** | InP | 44 | 220 | 0.77 |
| **D3** | InP | 72 | 220 | 0.22 |
| **D4** | InAs | 83 | 440 | 0.040 |
| **D5** | InAs | 83 | 110 | 0.026 |
| **D6** | InAs | 44 | 130 | 0.16 |

**Fig. S1.** The product of normal state resistance, $R_N$, and the nanowire-electrode contact area, $A$, is plotted as a function of the nanowire diameter for InAs (circle) and InP (triangle) nanowires.

**Fig. S2.** Differential conductance $dI/dV(V)$ traces for ($\phi$ = 72 nm, $L$ = 220 nm) at $B$ = 0 mT and different temperatures and $T$ = 1.65 (black), 1.27 (red), 1.10 (green), 0.85 (blue), 0.47 (cyan), and 0.02 (magenta) K. Left inset: SEM image of device **D3** (scale bar: 500 nm). Right inset: linear conductance $G(V_g)$ trace in the case of normal-state contacts ($B$ = 56 mT). The conductance fluctuations with $V_g$ are reproducible and have a rms amplitude of ~0.4 $e^2/h$.

Doh *et al.*, Fig. 1

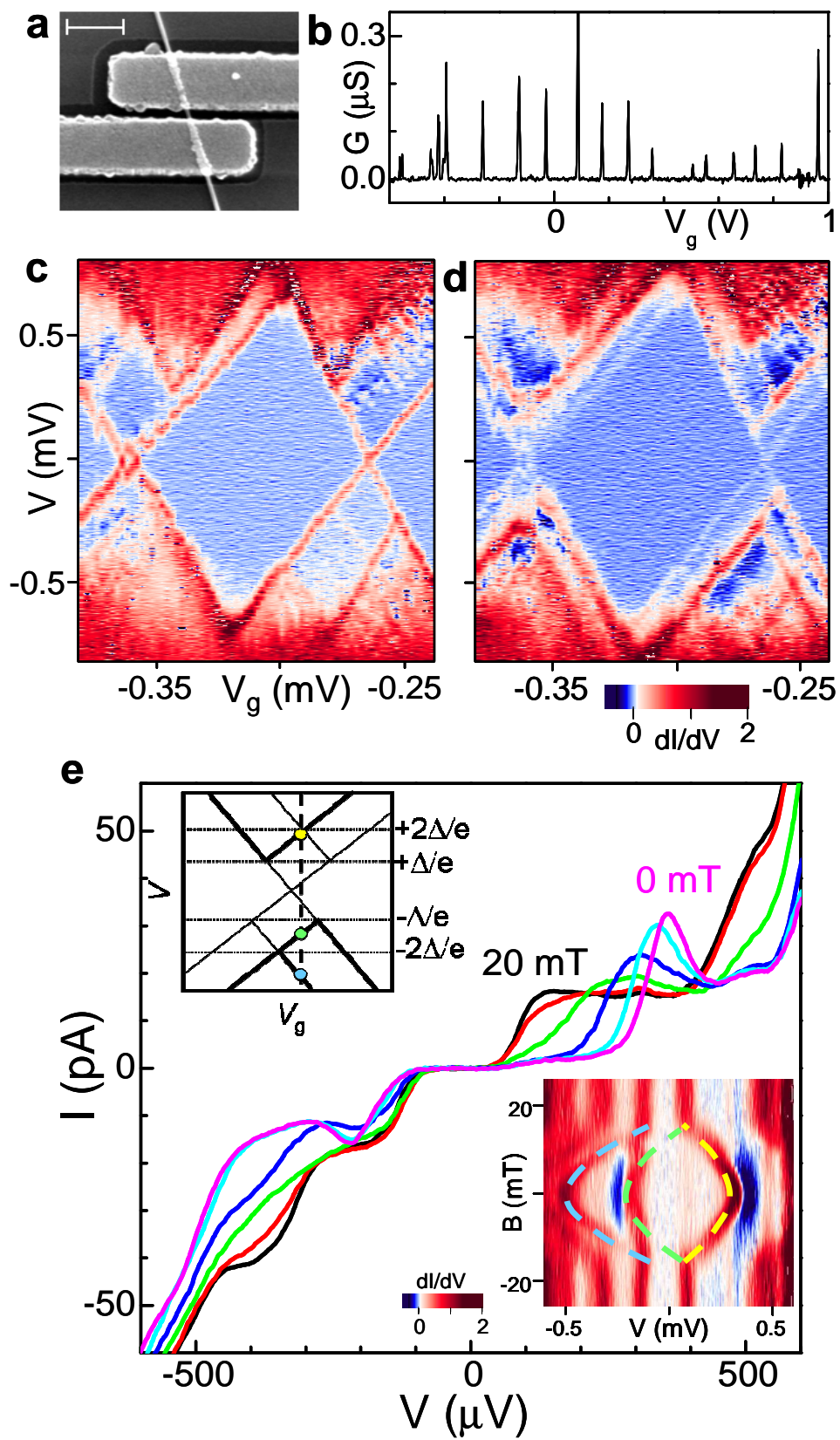

Doh *et al.*, Fig. 2

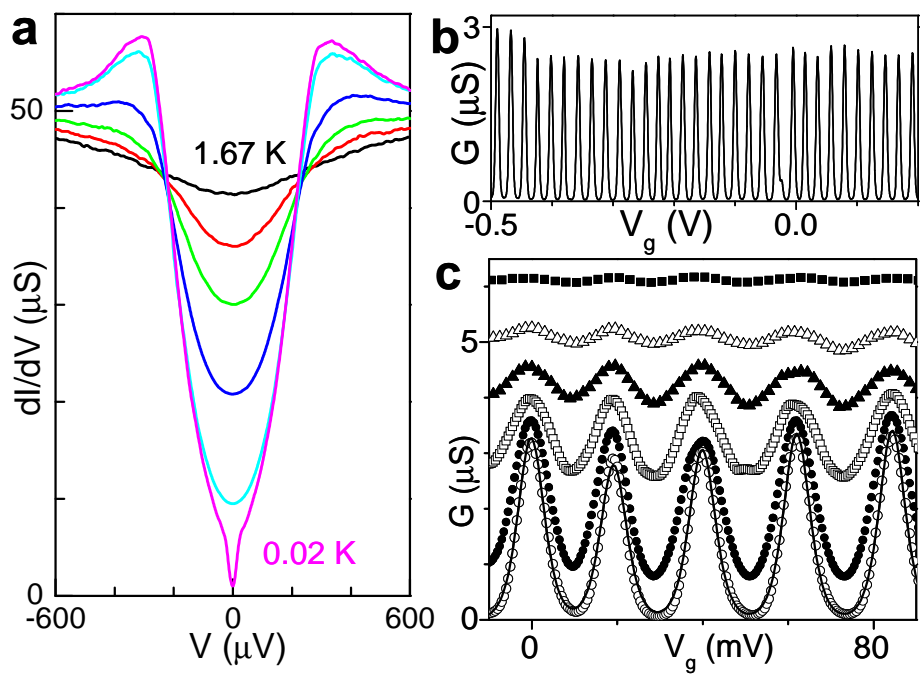

Doh *et al*., Fig. 3

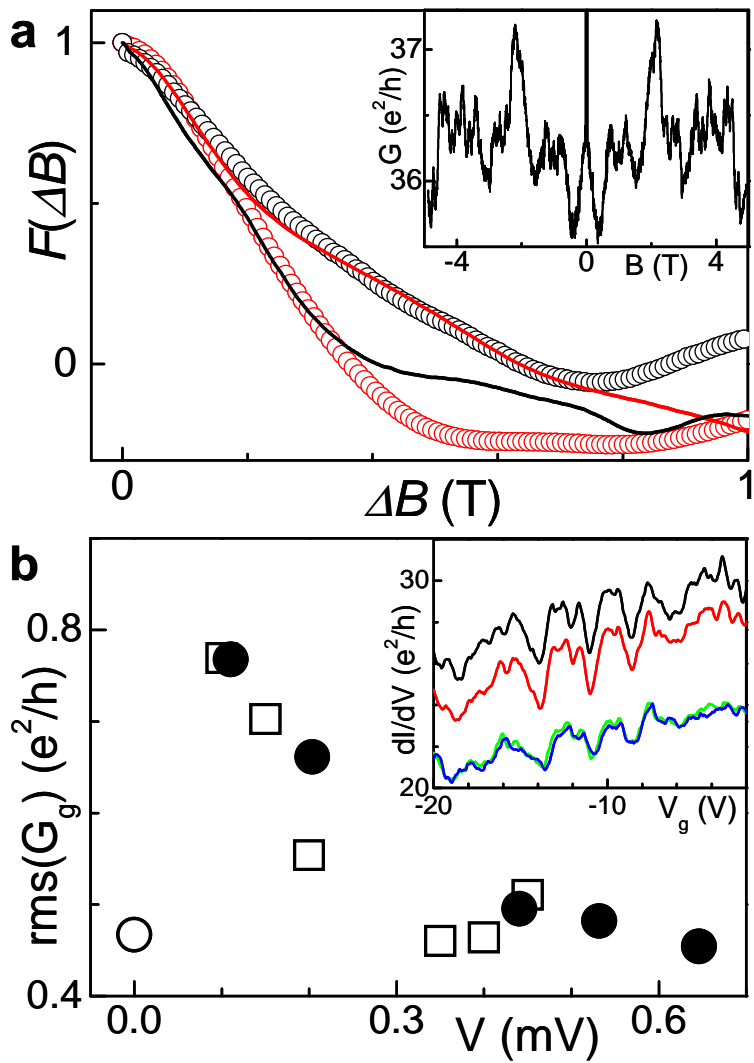

Doh *et al*., Supplementary Fig. S1

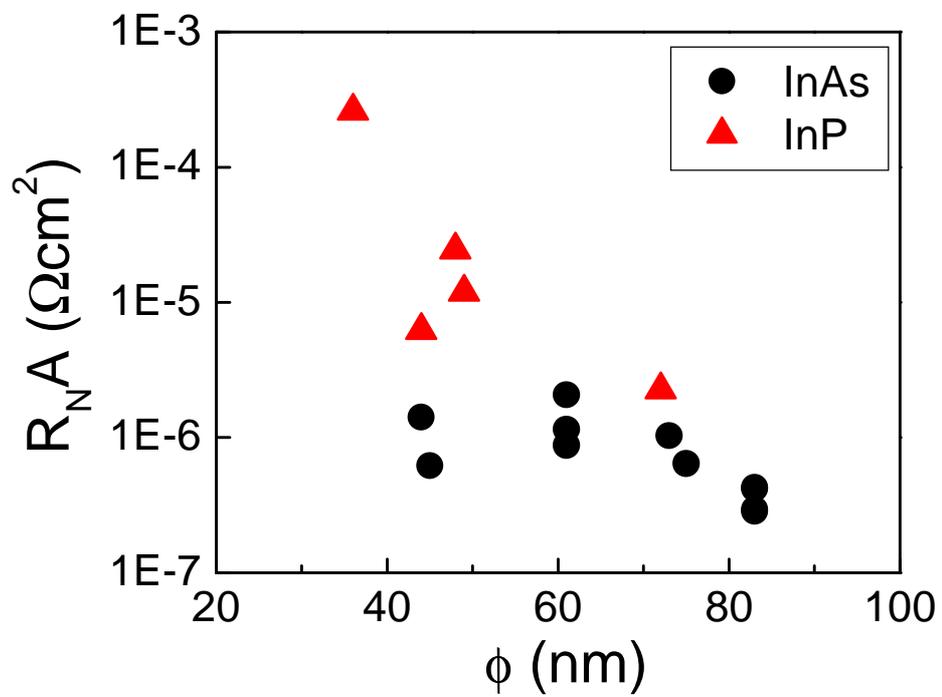

Doh *et al*., Supplementary Fig. S2

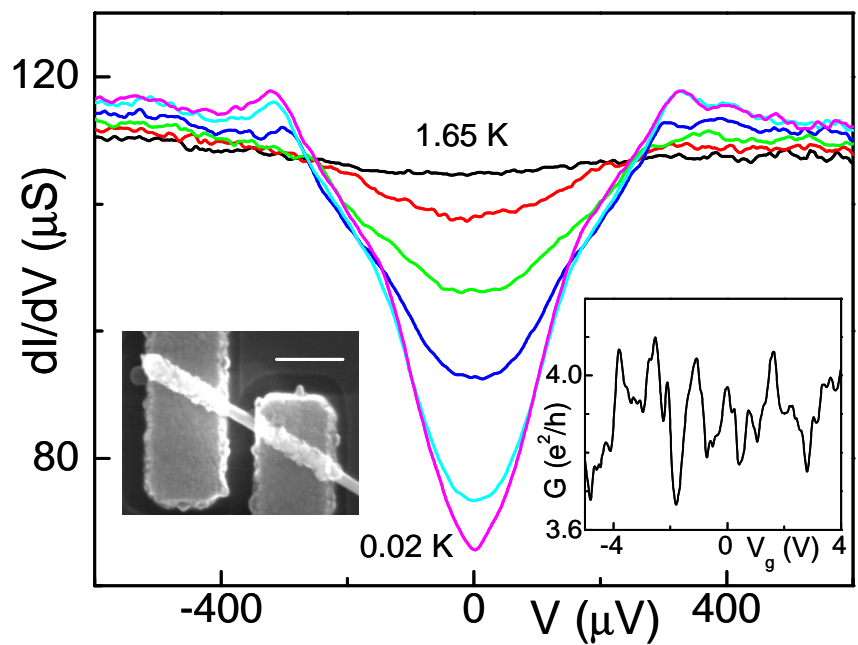